\documentclass[%
 preprint,
 amsmath,amssymb,
 aps, physrev,
]{revtex4-2}
\usepackage{xspace}
\usepackage{ragged2e}
\usepackage{multirow}
\usepackage{xcolor}
\usepackage{graphicx}
\usepackage{tabularx}

\begin{document}

%\preprint{}

\title{Digital-Analog concept for superconducting perceptron-like neural networks}

\author{Andrey E. Schegolev$^{1,2,*}$, Vsevolod I. Ruzhickiy$^{1,2}$, Georgy I. Gubochkin$^{2,3}$, Alexander S. Ionin$^3$, Ivan A. Nazhestkin$^3$, Mikhail Y. Fominskii$^4$, Lyudmila V. Filippenko$^4$, Igor I. Soloviev$^1$, Maxim V. Tereshonok$^2$, and Nikolay V. Klenov$^2$}

\affiliation{$^1$Skobeltsyn Institute of Nuclear Physics, Lomonosov Moscow State University, Moscow 119991, Russia}
\affiliation{$^2$Moscow Technical University of Communications and Informatics (MTUCI), 111024 Moscow, Russia}
\affiliation{$^3$Superconducting Quantum Computing Lab, Quantum Center, Skolkovo, Moscow Region, Russia}
\affiliation{$^4$Laboratory of superconducting devices for signal detection and processing, Kotelnikov Institute of Radioengineering and Electronics, Moscow, Russia}
\email{Contact author: tanuior@gmail.com}

\date{\today}

\begin{abstract}
A promising route to superconducting artificial neural networks is a hybrid digital-Analog architecture that combines digital single-flux-quantum (SFQ) communication with compact Analog nonlinear processing. The study focused on the dynamic conversion of a discrete signal passing through a digital-to-Analog-to-digital (DAD) converter, in which the role of the Analog cell was performed by a $\Sigma$-neuron with a nonlinear transfer function -- the basic cell of perceptron-like neural networks. Furthermore, the DAD converter, the elementary functional block of the hybrid architecture, combines a digital-to-analog converter (DAC) and an Analog-to-digital converter (ADC), and re-encodes the Analog $\Sigma$-neuron waveforms as an SFQ pulse sequence. Circuit-level simulations demonstrate how input values encoded by SFQ pulse trains are converted into analog signal levels, transformed by the $\Sigma$-neuron, and mapped back to pulse-based outputs. As a key experimental step, we fabricated and characterised a redesigned $\Sigma$-neuron and measured a sigmoid-like transfer characteristic suitable for activation-function implementation. The extracted response was incorporated into system-level simulations to assess the influence of realistic device parameters on the conversion process. We delineate the operating-range matching requirements for the DAC, neuron, and ADC blocks, supporting the feasibility of the proposed interface as a building block for perceptron-like superconducting neural networks with digital inputs and outputs.
Finally, we developed two perceptron networks, one using a mathematical sigmoid activation and the other the measured $\Sigma$-neuron transfer characteristic, which reached classification accuracies of $97.0\%$ and $91.9\%$, respectively, on the MNIST handwritten digit dataset.
\end{abstract}

\maketitle

\section{Introduction}
The rapid growth of artificial intelligence and machine learning has sharply increased the size of neural networks and the resources needed to learn and run them. In conventional digital hardware the energy cost is tied not only to arithmetic intensity but, crucially, to the von Neumann organisation, in which repeated data transfers between memory and processing units dominate power and latency. This has renewed interest in hardware that treats neural computation as the evolution of nonlinear dynamical elements and that uses event-based or hybrid digital--analog representations to reduce communication and precision overhead.

In this context superconducting electronics is an attractive route, offering ultra-low dissipation and high-speed operation at cryogenic temperatures \cite{segall2017synchronization, schneider2020synaptic, goteti2022superconducting, shainline2017optoelectronic, khan2022optoelectronic}. Several distinct superconducting neuromorphic directions have emerged. One keeps the computation entirely in the single-flux-quantum (SFQ) domain, where SFQ pulses act as spikes and the neuron is built from Josephson comparators and SFQ logic. High-throughput processors of this type, such as the SUSHI chip \cite{liu2023sushi}, reach about $32.4$~TSOPS/W at $42$~mW, but they synthesise the activation nonlinearity digitally. The so-called bioSFQ family \cite{semenov2022biosfq, semenov2023biosfq} instead bridges the digital and analog domains, storing synaptic information as a loop current transferred as a rate of SFQ pulses. Tunable synapses have also been realised with magnetic Josephson junctions, whose magnetic nanostructures set the junction critical current at attojoule energies \cite{schneider2020synaptic, schneider2018synapses}. Superconducting optoelectronic networks \cite{shainline2017optoelectronic, khan2022optoelectronic} combine few-photon emitters, nanowire single-photon detectors and optical waveguides for very large fan-out. Disordered Josephson-junction and superconducting-loop arrays \cite{goteti2022superconducting, goteti2021disordered} exploit the collective fluxon dynamics of many coupled loops. Among these technologies, rapid single-flux-quantum (RSFQ) logic \cite{mukhanov1987ultimate, krylov2023rapid} is particularly attractive, combining zero static resistance with picosecond switching and femtojoule-level energy per operation.

Against this background, the distinctive feature of our approach is that the activation element is an analog $\Sigma$-neuron whose transfer characteristic is fixed by measurements of a fabricated device and which provides the nonlinear map between digital SFQ interfaces through a DAC--$\Sigma$-neuron--ADC chain. Rather than forming the sigmoid in SFQ logic \cite{liu2023sushi}, distributing it across magnetic synapses \cite{schneider2018synapses}, or letting it emerge from disorder \cite{goteti2022superconducting, goteti2021disordered}, we keep robust SFQ pulse representations at the input and output and delegate only the nonlinearity to a compact analog cell. The novelty is therefore the constrained digital--analog--digital (DAD) signal path, with particular emphasis on the analog-to-digital return. The ADC analysis shows how the useful dynamic range of the $\Sigma$-neuron is mapped back to the SFQ domain, while the all-Josephson-junction DFF and TFF cells (Fig.~\ref{fig_FF}) make the digital parts compact and scalable. This distinguishes the work both from fully SFQ implementations, where the activation is synthesised digitally with relatively complex and energy-consuming circuits, and from fully analog superconducting circuits, where routing and programmability become difficult at scale. In our previous work, we discussed only the theoretical foundations of digital pulse control of an adiabatic $\Sigma$-neuron \cite{bastrakova2025digital}.

\begin{figure}
   \includegraphics[width=0.75\textwidth]{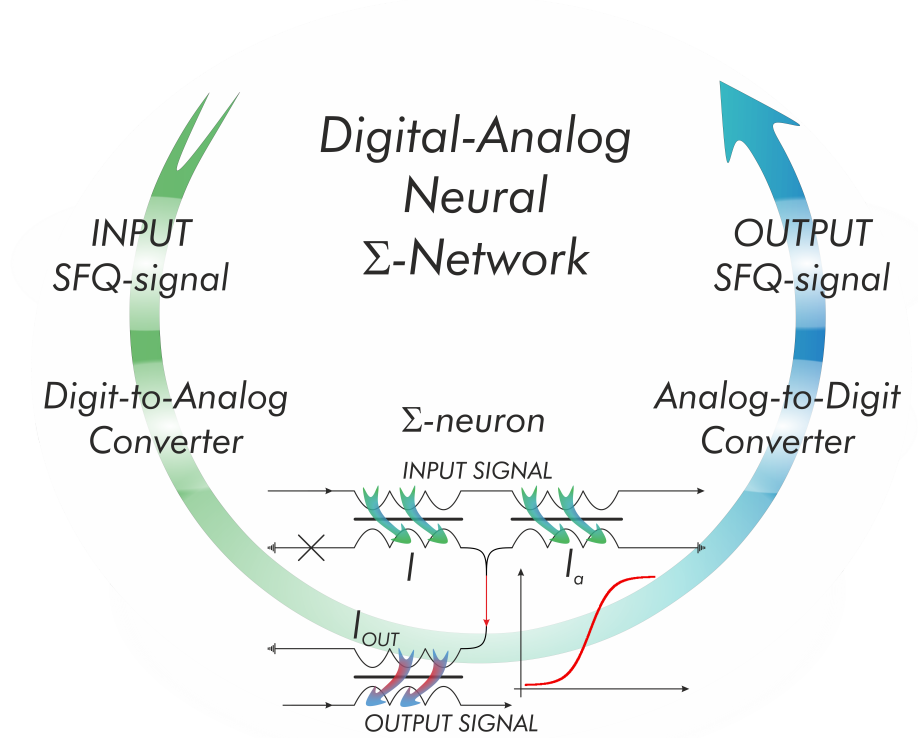}
   \caption{Illustration of concept of a digital-analog neural network using the example of a "DAC + $\Sigma$-neuron + ADC" combination as a functional part of it. Inset near the neuron's principal scheme demonstrates its transfer (activation) function
            }
    \label{fig1}
\end{figure}

In the proposed computational primitive  (Fig.~\ref{fig1}) an input symbol, encoded as an SFQ pulse word, is converted by a DAC \cite{nakanishi2012digital} into an analog level, processed by the analog $\Sigma$-neuron \cite{soloviev2018adiabatic}, and mapped back to an SFQ pulse sequence by an ADC \cite{mukhanov2004superconductor, gupta2011superconductor, radparvar2014superconductor}. The present work develops four elements of this primitive. First, a scalable SFQ-compatible DAC. Second, an experimentally characterised, redesigned $\Sigma$-neuron. Third, an ADC stage matched to the neuron output, which converts its current pulse into a calibrated SFQ count and then into a binary word with compact all-Josephson-junction cells. Fourth, a network-level test in which the measured neuron characteristic is used as the activation of a trainable perceptron.

This last step connects the work to hardware-aware training, in which a physical model of the hardware is placed inside the training loop so that the network learns with, rather than against, the device constraints \cite{Rasch2020}. The approach has been developed mainly for analog-memory and phase-change-memory accelerators \cite{Aguirre2024, Ambrogio2018, Joshi2020, Spoon2021, Burr2017}. Related schemes inject the measured device variation into the weights, treat memristor non-idealities collectively, and provide open analog-aware toolkits \cite{Boquet2020, Joksas2022, Rasch2021aihwkit}. For silicon analog networks, a physics-informed model recovers ideal-network accuracy despite strong non-idealities \cite{Filippeschi2026}. In all of these the constrained elements are the memory cells and converters, whereas the neuron activation is an idealised function. In this work we take the complementary step of making the neuron itself the constraint. Using the measured $\Sigma$-neuron characteristic as the hidden activation and holding the weights within the bounded range accessible to physical couplings, we train a single-hidden-layer perceptron to classify MNIST digits \cite{lecun1998mnist}. This network-level test (Sec.~\ref{Sec:MNIST}) shows that the fabricated $\Sigma$-neuron remains usable as an activation under realistic constraints and represents a first step toward physics-aware training for a superconducting element base.

%%%%%%%%%%%%%%%%%%%%%%%%%
\section{Model of digital-to-analog and analog-to-digital converters}
We now examine in greater detail the operation of superconducting digital‑to‑analog and analog‑to‑digital converters in conjunction with the $\Sigma$‑neuron cell. In other words we present a computational primitive implementing a measured, device-derived nonlinear map that can serve as an activation function inside network dynamics.

This analog neuron concept was first introduced a decade ago \cite{schegolev2016adiabatic} on the basis of adiabatic logic cells \cite{takeuchi2014reversible,  takeuchi2015thermodynamic} and was subsequently refined in further studies (e.g., \cite{soloviev2018adiabatic, schegolev2020learning, yamauchi2024dual}). 
The conditions under which such an adiabatic $\Sigma$-neuron can be driven directly by picosecond SFQ pulses, that is, controlled from the digital side, were recently established theoretically by an overlapping group of authors~\cite{bastrakova2025digital}. The present work builds on that digital-control result and addresses the complementary parts of the interface: the experimental characterisation of a fabricated $\Sigma$-neuron, and the analog-to-digital return that maps the neuron output back into an SFQ pulse count with compact inductor-free all-Josephson-junction cells.
The concept of digital processing in the proposed hybrid architecture is rooted in the propagation of soliton-like waves of current and voltage in Josephson junction chains \cite{ustinov1998solitons, wustmann2020reversible}. In RSFQ logic, a logical state ``1'' is encoded by the appearance of a short (picosecond-scale, "discrete" in terms of the analog part) voltage pulse across a Josephson junction, whereas a logical state ``0'' corresponds to the absence of such a pulse. These fluxon pulses travel along superconducting transmission lines without significant dispersion, behaving as topological solitons that retain their shape and energy. This soliton-based information carrier enables ultra-low-power, high-speed operation and provides a natural interface between digital and analog domains. The corresponding neuron, DAC and ADC schematics are presented in Figs.~\ref{fig1} (at the bottom) and \ref{fig_DAC_ADC}, respectively.

%% Sigma-neuron equations
All processes in discrete domain, as well as in an analog domain, related to the current flowing through a Josephson junction (JJ), determined by the following main equation: 
\begin{equation}\label{eq:mainJJ}
    I_{JJ} = C\frac{\Phi_0}{2\pi} \frac{d^2\varphi}{dt^2} + \frac{\Phi_0}{2\pi R_N}\frac{d\varphi}{dt} + I_C sin \varphi,
\end{equation}
where $\varphi$ is a phase across the Josephson junction, $I_C$, $R_N$ and $C$ are critical current, resistance in normal state and capacitance of Josephson junction and $\Phi_0$ is a magnetic flux quantum. This equation resembles, in form, the equation for a mathematical pendulum oscillating under the action of an external force in a viscous medium, where $\varphi$ is an rotation angle. A phase shift of 2$\pi$ is analogous to a full rotation of the pendulum about its pivot point. For digital RSFQ circuits, it is important that the phase $\varphi$ at the Josephson junction ‘rotates’ by 2$\pi$, as this generates a voltage pulse -- the basis for information transmission in such circuits. At the same time, for the analog part of the neural network — the $\Sigma$-neuron -- it is important that the phase $\varphi$ undergoes as small oscillations as possible, or at least makes a ‘full rotation’ as infrequently as possible, which will have a positive effect on the smoothness of its transfer function.

Based on this, we can write the motion equation for a $\Sigma$-neuron (the result of solving the phase equations and Kirchhoff’s laws for the circuit shown in the bottom of the figure \ref{fig1}):
\begin{equation}\label{eq:mainSigma}
    \beta \ddot{\varphi} + \alpha \dot{\varphi} + sin \varphi = b \cdot \varphi_{in}(t) - a \cdot \varphi,
\end{equation}
where $\beta$ (normalised capacitance, $\beta = 2\pi I_C C / \Phi_0$) and $\alpha$ (damping factor, $\alpha$ = $\sqrt{\Phi_0 / (2\pi I_C R_N^2 C)}$) are equal to 1, $\varphi(t)$ -- is a Josephson junction phase,  $b = \frac{l_a + 2l_{out}}{2 \cdot (l l_a + l_{out}(l + l_a))}$ and $a = \frac{l_a + l_{out}}{l l_a + l_{out}(l + l_a)}$, $l$, $l_a$ and $l_{out}$ are $\Sigma$-neuron's inductances of Josephson, non-Josephson and output branches respectively (the symbols are shown in the fig. \ref{fig1}). Here $\varphi_{in}(t)$ is the converted digital signal from the DAC supplied to the input of the $\Sigma$-neuron. Dot notation means a time derivative. Solution of this equation substituted to the output $\Sigma$-current as follows:
 \begin{equation}\label{eq:outSigma}
     i_{out}(t) = \frac{\varphi_{in}(t) - 2 l_a \cdot (b \cdot \varphi_{in}(t) - a \cdot \varphi(t))}{2 \cdot (l_a + l_{out})}.
 \end{equation}
The current $i_{\rm out}(t)$ is the analog signal delivered to the ADC. In the block-level simulations used below, the interface between the neuron and the ADC is represented by
\begin{equation}\label{eq:ADCinput}
    i_{\rm ADC}(t)=i_0+K_{\rm out} i_{\rm out}(t),
\end{equation}
where $i_0$ fixes the ADC bias point and $K_{\rm out}$ is an effective coupling or gain coefficient of the output interface. In an integrated implementation, this coefficient will be determined by the particular mutual or galvanic coupling geometry, by attenuation in the interconnect, and by the ADC input impedance. In the present work it is used as a calibrated matching parameter chosen so that the useful range of the $\Sigma$-neuron output falls within the linear counting range of the ADC shown in Figures \ref{fig_ADC_Demo} and \ref{fig_pulses}.

\subsection{D and T flip-flop operation}
\begin{figure}
    \includegraphics[width=0.95\linewidth]{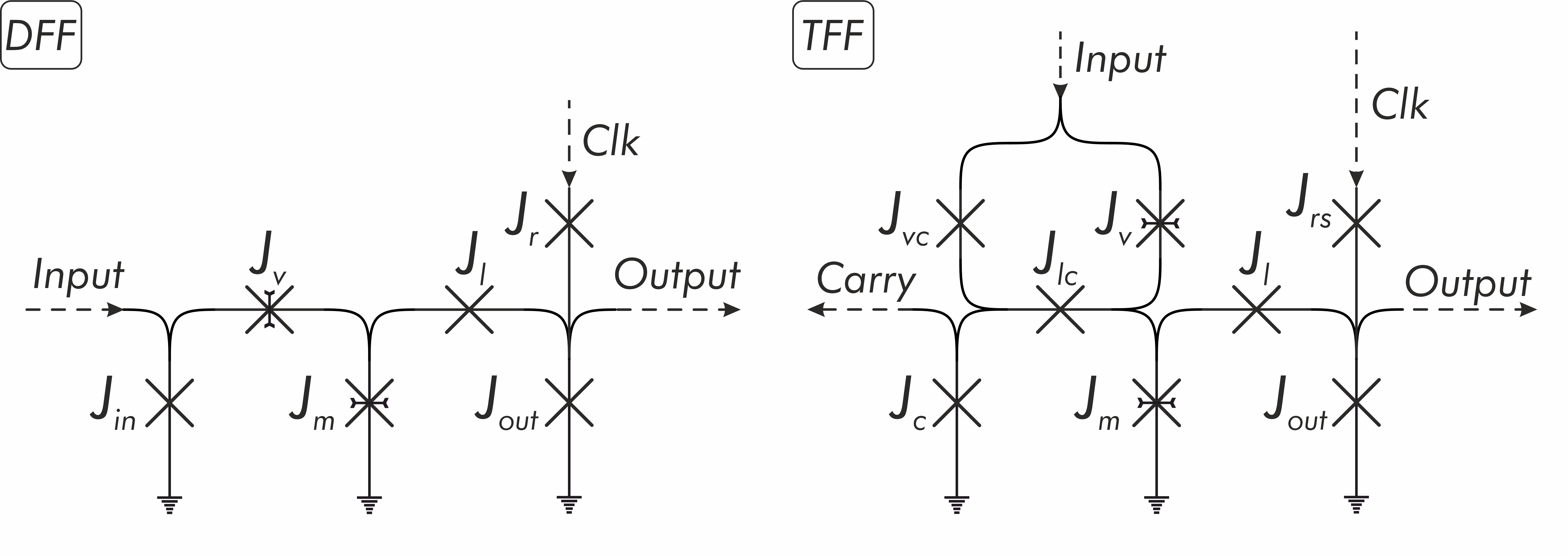}
    \caption{\label{fig_FF} Inductor-free all-Josephson-junction storage cells used in the digital parts of the DAD interface. Left: D flip-flop (DFF) used in the DAC input register as a clocked gate for the weighted SQUID conversion elements. Right: T flip-flop (TFF) used in the ADC output counter. Each input SFQ pulse toggles the internal state of the TFF. An overflow pulse is generated on the CARRY line and drives the next counter stage, while the clock/read pulse transfers the stored bit to the OUT line and resets the stage for the next conversion cycle. Two crossed lines (a cross) denote ordinary Josephson 0-junctions, while bistable junctions ($J_m$ and $J_v$) are marked with an additional line crossing the cross
            }    
\end{figure}

The key digital components of the DAC and ADC used in this work are the inductorless D flip-flop (DFF) and T flip-flop (TFF), whose circuit diagrams are shown in Fig.~\ref{fig_FF}. Such cells have been part of the digital superconducting-electronics library for more than two decades \cite{Polonsky1994, Bairamkulov2024, Matsumoto2024}. The inductor-free implementation used here was first proposed by the present authors in \cite{soloviev2021superconducting}. Both flip-flops contain bistable Josephson junctions, denoted $J_m$ and $J_v$ in Fig.~\ref{fig_FF}, whose current-phase relation contains in such configuration contain only the second harmonic, $I_{bsJJ}\approx I_c\sin(2\varphi)$. Their dynamics follow Eq.~(\ref{eq:mainJJ}) with $I_c\sin(2\varphi)$ in place of $I_c\sin\varphi$, so such a junction has stable states at $\varphi=\pi n$, in contrast to the $\varphi=2\pi n$ states of an ordinary 0-junction. In both cells the stored logic state is encoded in the phase of the junction $J_m$.

The DFF (Fig.~\ref{fig_FF}) operates as follows. An input (data) pulse on the \textit{Input line} switches the flip-flop from state ``0'' to state ``1'', which corresponds to a change of the phase across $J_m$ from $0$ to $\pi$. A subsequent pulse on the \textit{Clk line} resets the state from ``1'' back to ``0''. In state ``1'' the current leaking into the \textit{Output line}, which feeds a SQUID within the DAC block, is larger than in state ``0''. Consequently, the \textit{Output line} delivers a current pulse whose duration is set by the delay between the data pulse and the subsequent read pulse.

In the TFF, when $J_m$ changes its state the circulating currents in the adjacent loops change as well. The state (the phase on $J_m$) is toggled by pulses on the \textit{Input line}: in state ``1'' ($\varphi_m\approx\pi$) an input pulse also switches the $J_{cout}$ and emits a pulse on the \textit{Carry line}. The state is read out by a comparator formed by $J_{rs}$ and $J_{out}$: when a \textit{Clk} pulse arrives, whichever junction is closer to its critical current switches. In state ``1'' the current in the loop $J_m$--$J_l$--$J_{out}$ brings $J_{out}$ closer to its critical value, and its switching emits a pulse on the \textit{Output line}. The same \textit{Clk} pulse also resets the flip-flop, the phase on $J_m$ returning from $\pi$ to $2\pi$ (state ``0'').

\subsection{DAC operation}
\begin{figure*}
 \includegraphics[width=0.75\textwidth]{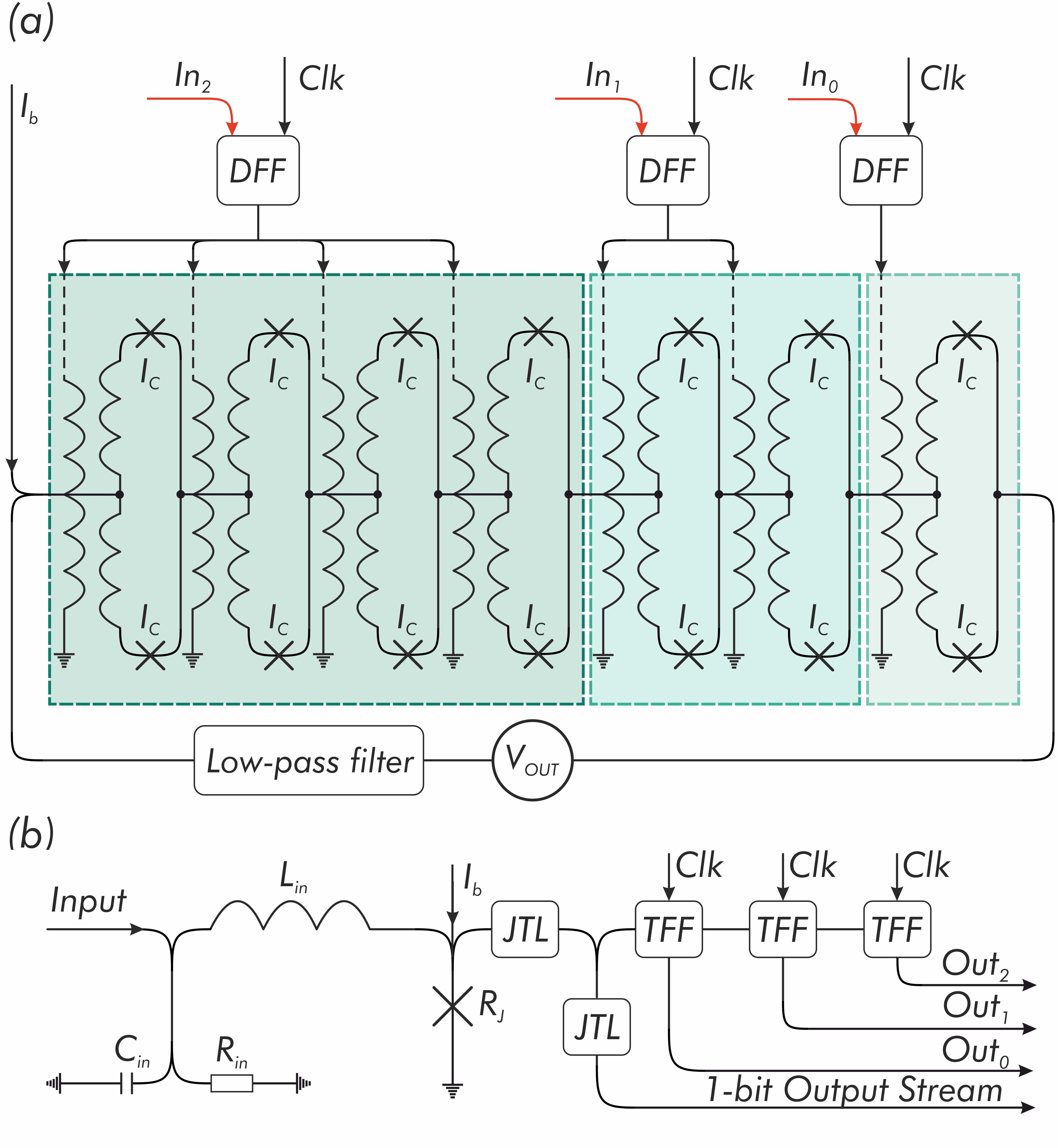}
 \caption{\label{fig_DAC_ADC} Schematic representation of (a) a digital-to-analog and (b) analog-to-digital converter. $In_{0,1,2}$ and $Out_{0,1,2}$ correspond to the 0-, 1- and 2-bit streams accordingly. All converters here have a 3-bit resolution
         }
\end{figure*}

The DAC consists of an array of D flip-flops at the input, whose number is determined by the required bit depth, which act as gates: they pass clock pulses to subsequent stages of the register if the DFF is in the "1" state, and do not pass them if the DFF is in the "0" state. Two-junction SQUIDs are used as the conversion elements (see fig. \ref{fig_DAC_ADC}a). SQUIDs in state the ''1'' switch to the resistive state and generate an output voltage $V$ during the clock signal, while SQUIDs in state ''0'' remain in the superconducting state (with zero output voltage). The DAC output is given by the sum of the voltage contributions of all active SQUIDs, which can be smoothed or averaged using a corresponding filter.
The number of SQUIDs in DAC assigned to each trigger corresponds to the number of digits: the zero digit corresponds to one SQUID (output voltage $0$ or $V$), the first digit corresponds to two SQUIDs (output voltage $0$ or $2V$), the second to four SQUIDs (output voltage $0$ or $4V$), and so on. The voltage read from these SQUIDs corresponds to a digital signal in binary representation.

In this study we used the full $\{0,1,2,3,4,5,6,7\}$ input set, which spans the entire information range representable by a 3-bit DAC and ADC ($2^3=8$ levels), so that the chain is exercised over all output codes rather than a partial subset. The conversion of the digital sequence into an analog representation is shown in Fig. \ref{fig_pulses}a. The 3-bit DAC output signal was then applied to the input of the $\Sigma$-neuron. The input signal level determines the operating point on the neuron transfer characteristic. The corresponding $\Sigma$-neuron output is shown in Fig.~\ref{fig_pulses}a. Clearly, the $\Sigma$-neuron converts external signals with different levels in different ways. After nonlinear transformation in the $\Sigma$-neuron ($\Sigma$-neuron conversion), the analog signal is delivered to the 3-bit ADC input. Here, signal transmission and processing were simulated for the individual functional blocks (DAC, $\Sigma$-neuron, ADC) separately -- the full DAC--$\Sigma$-neuron--ADC chain was not co-simulated as a single self-consistent circuit. We focused here on looking at individual functional blocks, which can later be combined to form more complex circuits.

\subsection{ADC operation}
\begin{figure*}
 \includegraphics[width=0.95\textwidth]{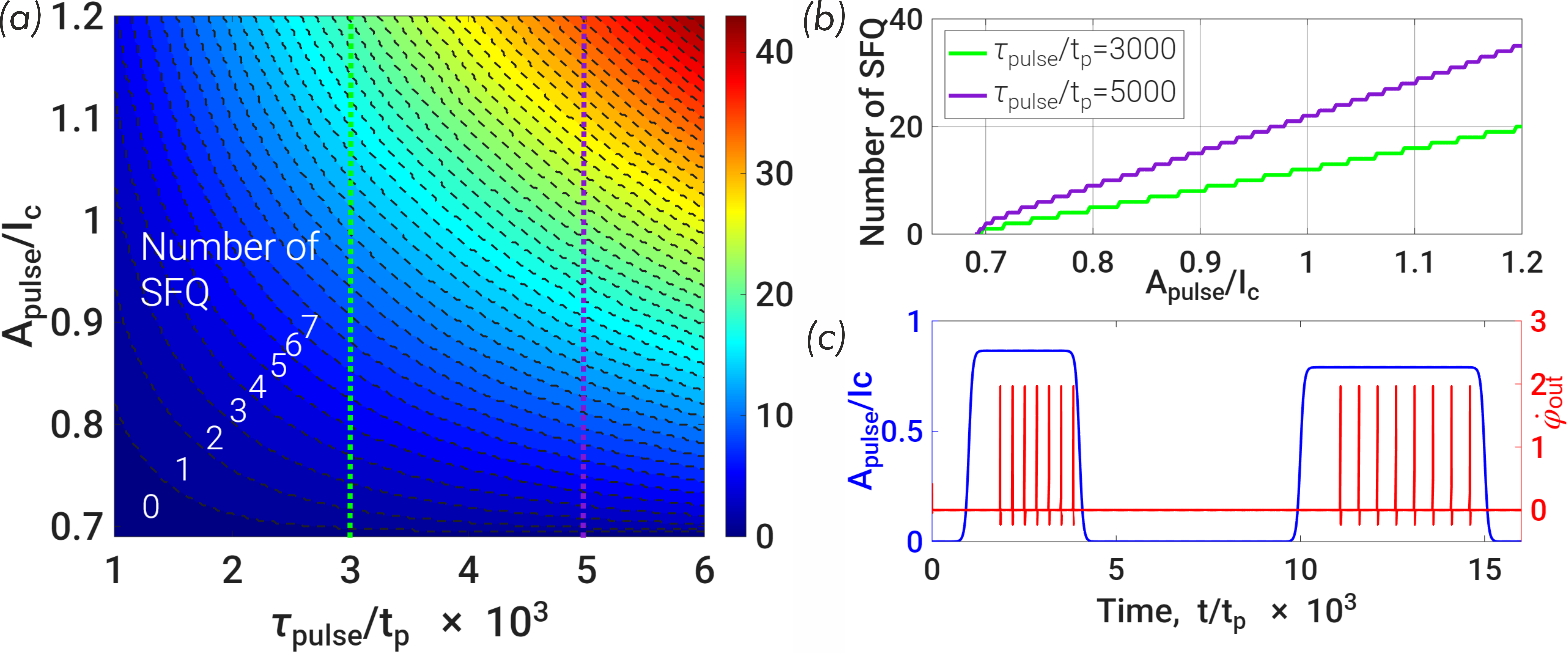}
 \caption{\label{fig_ADC_Demo} Calibration and time-domain operation of the ADC front end. (a) Simulated number $N_{\rm SFQ}$ of SFQ pulses delivered to the JTL as a function of the input-pulse amplitude $A_{\rm pulse}/I_c$ and duration $\tau_{\rm pulse}/t_p$. The dashed contours indicate constant pulse count and define the available counting range. (b) Cross-sections of the calibration map at fixed durations $\tau_{\rm pulse}/t_p=3000$ and $5000$, showing an approximately linear amplitude-to-count conversion after threshold-offset correction. (c) Representative time traces of the ADC input current pulses, shown in blue, and the phase derivative $\dot{\varphi}_{\rm out}$ at the last JTL junction, shown in red. Each sharp peak in $\dot{\varphi}_{\rm out}$ corresponds to an emitted SFQ pulse
         }
\end{figure*}

% front end + counting principle
At the ADC front end a rectangular current step drives a Josephson junction, shunted within an $RC$ input network, into its resistive state; while the junction is resistive it emits SFQ pulses into the transmission line, so the pulse count depends on both the amplitude and the duration of the step. Figure~\ref{fig_ADC_Demo}a maps this count as a function of amplitude and duration. When the duration is fixed, equal increments of the amplitude add one pulse to the count (Fig.~\ref{fig_ADC_Demo}b), apart from the first one or two pulses near threshold, which are removed by a constant offset in the read-out. For the 3-bit converter used here and a duration $\tau_{\rm pulse}/t_p = 3000$, a single linear fit passes through the seven discrete count steps that span the full dynamic range. Representative time traces of the phase derivative $\dot{\varphi}_{\rm OUT}(t)$ at the last JTL junction are shown in Fig.~\ref{fig_ADC_Demo}c.

\begin{figure*}
 \includegraphics[width=0.95\textwidth]{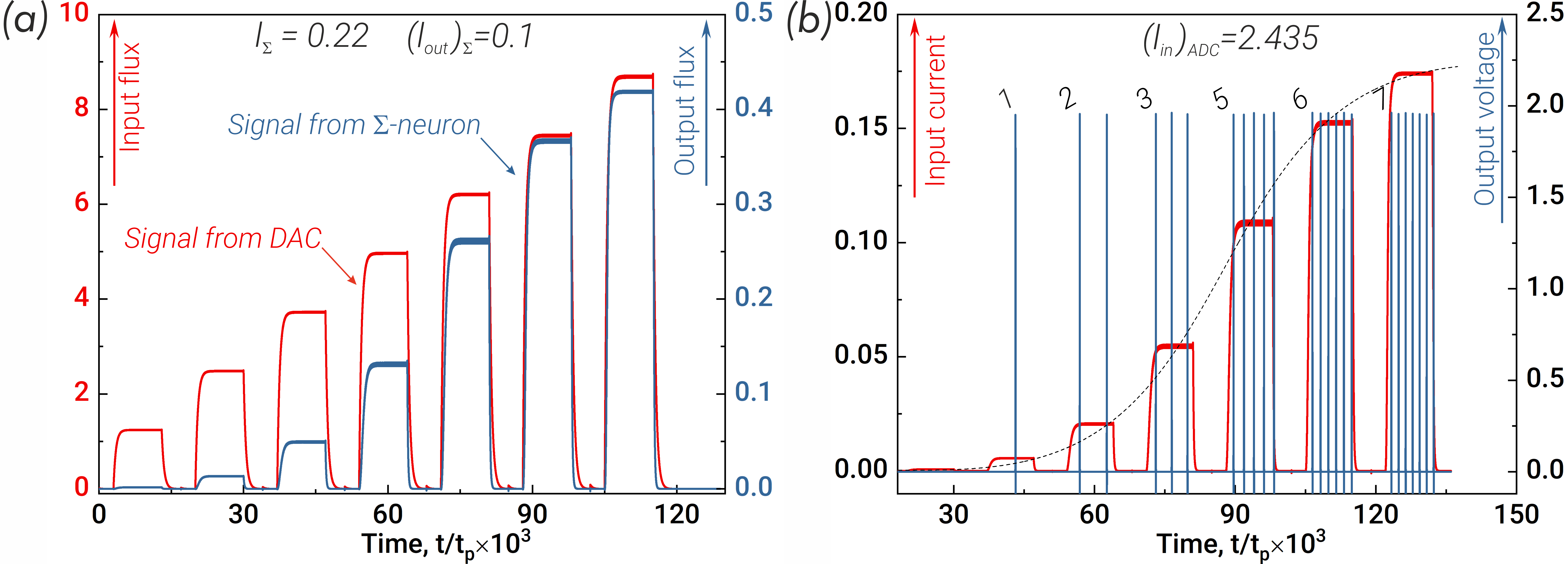}
 \caption{\label{fig_pulses} (a) Modelling the conversion of the $\{0,1,2,3,4,5,6,7\}$ SFQ signal by 3-bit DAC (red curves) and its processing by theoretically best $\Sigma$-neuron (dark-blue curves). (b) Simulation 3-bit ADC voltage response (dark-blue curves) after conversion of the signal, generated by $\Sigma$-neuron (red curves). Also, the figure shows the transfer function of the theoretically best $\Sigma$-neuron as a black dashed line to facilitate the interpretation of the ADC input signal. Parameters of the $\Sigma$-neuron are: $l = 0.22$, $l_{out} = 0.1$, $l_a = 1 + l$. Parameters of the DAC are: $\tau_{LPF}=10^4$, $i_b=1.8$. Parameters of the ADC are: $L_{in}/L_0=2.435$, $K_{\rm out}=0.41$, $(\omega_p)_{ADC}/(\omega_p)_{\Sigma} = 4.5$, for integrator $R_{int}/R_N = 0.1$, $L_{int}/L_0 = 40$, $R_J/R_N = 0.2$, $i_{b0} = 0.75$, for JTL $L_{JTL}/L_0 = 3$, $i_{bJTL}=0.75$, the length is equal to $10$. The values of all capacities were set equal to 1 ($\beta = 2\pi I_{C0} C / \Phi_0$ = 1), normalised inductance was set as $L_0=\Phi_0 / (2\pi I_C)$ and normalised time was set as $\sqrt{\Phi_0 C / (2\pi I_{C0})}$
         }
\end{figure*}

% coupling the neuron output into the counting window
To use this counter as an analog-to-digital converter for the neuron, the $\Sigma$-neuron output must fall within its linear counting window. The neuron output enters the ADC as $i_{\rm ADC}=i_0+K_{\rm out}\,i_{\rm out}$ (Eq.~(\ref{eq:ADCinput})) with a reference bias $I_0/I_c = 0.69$, a pulse of amplitude $A_{\rm pulse}/I_c = 0.025\,n$ then yields $n$ output pulses. The coefficient $K_{\rm out}$ is chosen so that the useful range of the neuron output is mapped onto this linear range. For a direct galvanic connection with matched normalisation it reduces to an effective inverse inductance. In general it depends on the interface circuitry, the signal attenuation, and the normalisation conventions (the values used are given in the captions of Figs.~\ref{fig_pulses} and \ref{fig_pulses_experiment}).

% read-out into an N-bit word (cross-reference, no re-description)
The SFQ train is then accumulated by the chain of $N$ toggle flip-flops introduced above (Fig.~\ref{fig_FF}), which forms a binary ripple counter: each pulse toggles the first stage, the carries propagate to the following stages, and a clock pulse latches the result as an $N$-bit word. The governing equations of the all-JJ TFF are given below in Eqs.~(\ref{EqADC_TFF})--(\ref{EqADC_TFF_B}).

% ADC equations fullset
To place the ADC model on the same footing as the $\Sigma$-neuron equations~\ref{eq:mainJJ}--\ref{eq:outSigma}, we now give its governing equations. Although the ADC was simulated as a single circuit, it is convenient to view it as three blocks: the signal-detection scheme, the transmission line, and the TFF. The detection scheme is described by
\begin{equation}\label{EqADC_Det_1}
\ddot{\varphi}_{in} = i_{in} - \alpha_{in} \dot{\varphi}_{in} + \frac{\varphi_{in} - \varphi_d}{l_{in}},
\end{equation}
\begin{equation}\label{EqADC_Det_2}
\ddot{\varphi}_{d} = - \alpha_{d} \dot{\varphi}_{d} - \frac{\varphi_{in} - \varphi_d}{l_{in}} - i_{toJTL},
\end{equation}
where $\varphi_{in}$ and $\varphi_{d}$ are the node phases at the circuit input and at the detecting Josephson junction. Since $C_{in}=C_d$, no coefficient appears in front of the second-derivative terms; $\alpha_{in} = \frac{\omega_p \Phi_0}{2\pi I_c}\frac{1}{R_{in}}$ and $\alpha_{d} = \alpha_{in} \frac{R_{in}}{R_d}$ are the damping coefficients (the same damping parameter as in Eq.~\ref{eq:mainSigma}, written through the plasma frequency $\omega_p=\sqrt{2\pi I_c/(\Phi_0 C)}$); $l_{in}$ is the normalised inductance of the input signal converter; and $i_{toJTL}$ is the current component leaking into the transmission line.

The transmission line can be either a conventional Josephson transmission line (JTL) or a line built solely from Josephson junctions (all-JJTL). The equations below are written for the JTL and the all-JJTL, respectively, for an arbitrary node $k$, using the same normalisation to the critical current:
\begin{equation}\label{EqADC_JTL}
\ddot{\varphi}_{k} = i_{b} - \alpha \dot{\varphi}_{k} - \sin\varphi_k + \frac{\varphi_{k-1} - 2\varphi_k + \varphi_{k+1}}{l_{JTL}},
\end{equation}
\begin{equation}\label{EqADC_AllJJTL}
3\ddot{\varphi}_{k} - \ddot{\varphi}_{k-1} - \ddot{\varphi}_{k+1} = \\
i_{b} - \alpha \dot{\varphi}_{k} - \sin\varphi_k
+ \alpha (\dot{\varphi}_{k-1} - 2\dot{\varphi}_{k} + \dot{\varphi}_{k+1})
+ A_{con}\sin(\varphi_{k-1} - \varphi_k) + A_{con}\sin(\varphi_{k+1} - \varphi_k),
\end{equation}
where $i_{b}$ is the bias current, which acts as the driving force for soliton motion; $\alpha = \frac{\omega_p \Phi_0}{2\pi I_c}\frac{1}{R_N}$ is the damping parameter; $l_{JTL}$ is the normalised inductance of a JTL cell; in the all-JJTL all junctions share the same capacitance and damping; and $A_{con}$ is the normalised critical current of the connecting Josephson junction.

We denote the components of $\boldsymbol{\varphi}$ by $\varphi_i$, with $i \in \{m,\,\mathrm{Input},\,\mathrm{cout},\,\mathrm{out},\,\mathrm{Clk}\}$. For the TFF it is convenient to write the system of equations in matrix form for the vector of phases,
\begin{equation}\label{EqADC_TFF}
\hat{\mathbf{M}} \ddot{\boldsymbol{\varphi}} + \hat{\mathbf{M}} \dot{\boldsymbol{\varphi}} = \mathbf{B},
\end{equation}
where $\hat{\mathbf{M}}$ is the mass (capacitance) matrix. In our case all junctions have equal capacitance and damping parameters $\alpha$, so the matrix multiplying the first derivatives (damping matrix) coincides with the capacitance matrix:

\begin{equation}\label{EqADC_TFF_M}
\hat{\mathbf{M}} =
\begin{pmatrix}
 4 & -1 & -1 & -1 &  0 \\
-1 &  2 & -1 &  0 &  0 \\
-1 & -1 &  3 &  0 &  0 \\
-1 &  0 &  0 &  3 & -1 \\
 0 &  0 &  0 & -1 &  1
\end{pmatrix}.
\end{equation}
The vector $\mathbf{B}$ collects the remaining terms; the numerical coefficients correspond to the specific circuit parameters used here:
\begin{widetext}
\begin{equation}\label{EqADC_TFF_B}
\mathbf{B} =
\begin{pmatrix}
 i_b^m - B_m \sin(2\varphi_m) + B_v \sin\!\big(2(\varphi_{\mathrm{Input}} - \varphi_m)\big) + A_{lc}\sin(\varphi_{\mathrm{cout}} - \varphi_m) + A_l\sin(\varphi_{\mathrm{out}} - \varphi_m)\\[3pt]
 i_{\mathrm{Input}} - B_v\sin\!\big(2(\varphi_{\mathrm{Input}} - \varphi_m)\big) - A_{vc}\sin(\varphi_{\mathrm{Input}} - \varphi_{\mathrm{cout}}) \\[3pt]
 -i_{\mathrm{Carry}} + i_b^c - A_c\sin(\varphi_{\mathrm{cout}}) + A_{vc}\sin(\varphi_{\mathrm{Input}} - \varphi_{\mathrm{cout}}) - A_{lc}\sin(\varphi_{\mathrm{cout}} - \varphi_m)\\[3pt]
 -i_{\mathrm{Output}} + i_b^{out} - A_{out}\sin(\varphi_{\mathrm{out}}) - A_l\sin(\varphi_{\mathrm{out}} - \varphi_m) - A_{rs}\sin(\varphi_{\mathrm{Clk}} - \varphi_{\mathrm{out}}) \\[3pt]
 i_{\mathrm{Clk}} + A_{rs}\sin(\varphi_{\mathrm{Clk}} - \varphi_{\mathrm{out}})
\end{pmatrix}.
\end{equation}
\end{widetext}
It should be noted that the interface currents may also contain first- and second-order derivatives, which must then be included in the coefficient matrix. The normalised bias currents supplied to junctions $J_m$, $J_c$, $J_{out}$ are $i_b^m = 0.65$, $i_b^c = 0.6$, $i_b^{out} = 1.1$, respectively. For the bistable junctions, the normalised critical currents are $B_m = 0.7$ and  $B_v = 0.1$. The 0‑junctions, in turn, have normalised critical currents given by $A_c = 1$, $A_{out} = 1.5$, $A_{vc} = 0.1$, $A_{lc} = 0.4$, $A_{l} = 0.5$ and $A_{rs} = 0.1$.

The result of processing information through the chain "DAC -- $\Sigma$-neuron -- ADC" is shown in the Fig. \ref{fig_pulses}b. The number of pulses at the3-bit ADC output is denoted by the corresponding number next to the groups of output pulses. As shown, only the last three rectangular pulses correspond to the original value encoded by the DAC.
The initial digit $\{1\}$ (as well as $\{0\}$) is converted to $0$ (no output pulses), $\{2\}$ -- to $1$, $\{3\}$ -- to $2$, $\{4\}$ -- to $3$, $\{5\}$ -- to $5$, $\{6\}$ -- to $6$ and $\{7\}$ is converted to $7$ output pulses. The transfer function of a $\Sigma$-neuron is shown in the background of the Fig. \ref{fig_pulses}b as a black dashed line to facilitate the interpretation of the ADC input signal.

Simulation of the ``DAC -- $\Sigma$-neuron -- ADC'' chain shows that the per-symbol processing time reaches the order of tens of nanoseconds, set by the adiabatic dynamics of the $\Sigma$-neuron. The corresponding timing and energy estimates are discussed in the ``Discussion'' section.

\begin{figure*}
    \includegraphics[width=0.75\linewidth]{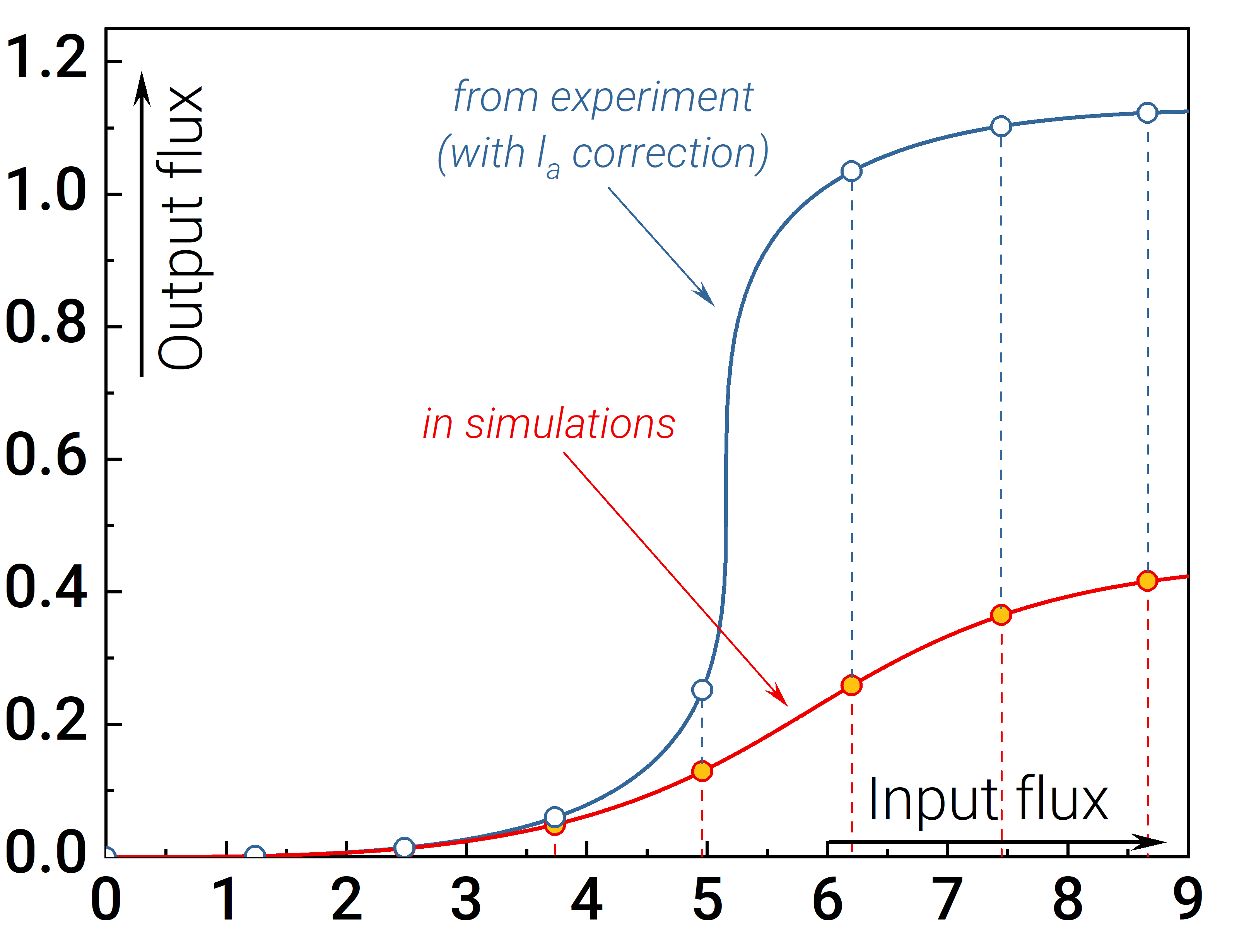}
    \caption{\label{fig_sigma_transfer} Transfer characteristics of a $\Sigma$-neuron in simulation (red curve, model-optimized) and from the analysis of obtained experimental data (dark-blue curve). The dots on the red curve (yellow inside) indicate the values of the input signal of a $\Sigma$-neuron in simulation, converted by DAC, while dots on the dark-blue curve (empty inside) demonstrate the possible nonlinearity of the same input signal conversion in the experiment. Parameters for the red curve: $l=0.22$, $l_{out}=0.1$, $l_a = 1 + l$. Parameters for the dark-blue curve: $l=0.65$, $l_{out}=0.465$, $l_a = 1.65$
            }
\end{figure*}

It is also important to comment on the results obtained at the ADC output. Figure \ref{fig_sigma_transfer} clearly shows that the values of the input flux in simulation (signal converted by the DAC) at levels $\{5\}$, $\{6\}$ and $\{7\}$ lie on the linear section of the $\Sigma$-neuron transfer characteristic (red curve), which is reflected in the direct correspondence between the initial signal and the final signal at the ADC output (Fig. \ref{fig_pulses}): $\{5\}$ was converted to five voltage pulses, $\{6\}$ was converted to six voltage pulses, $\{7\}$ was converted to seven voltage pulses. However, digits from $\{2\}$ to $\{4\}$ lie in the nonlinear region and are observed as one, two and three voltage pulses at the output of the chain, respectively. In the case of $\{1\}$, which falls on the practically zero section of the transfer characteristic, no output voltage pulses are observed. Thus, this simple example demonstrates in simulation the operation of the 3-bit DAC and ADC, and the $\Sigma$-neuron as a processing unit, as well as the features of the neuron's transfer characteristic: zero, nonlinear and linear sections.

%%%%%%%%%%%%%%%%%%%%%%%%%%%
\section{Experimental investigations of the $\Sigma$-neuron}
\begin{figure*}
    \includegraphics[width=0.95\linewidth]{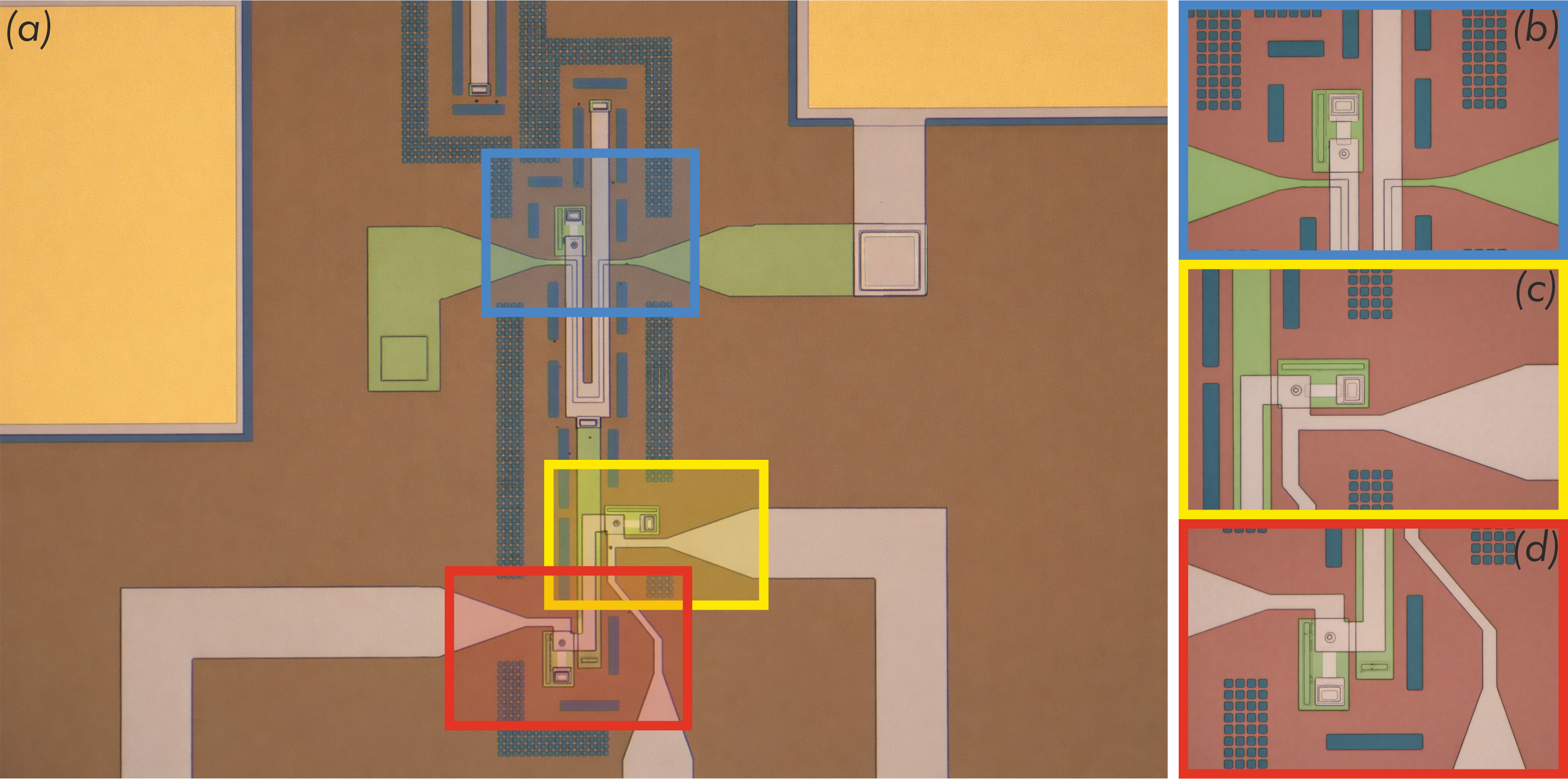}
    \caption{\label{fig_experiment_photo} Micrograph of the (a) fabricated $\Sigma$-neuron with signal line and readout SQUID, (b) zoomed area of the $\Sigma$-neuron Josephson junction and (c, d) zoomed areas for readout SQUID parts
            }    
\end{figure*}

\begin{figure*}
    \includegraphics[width=0.95\linewidth]{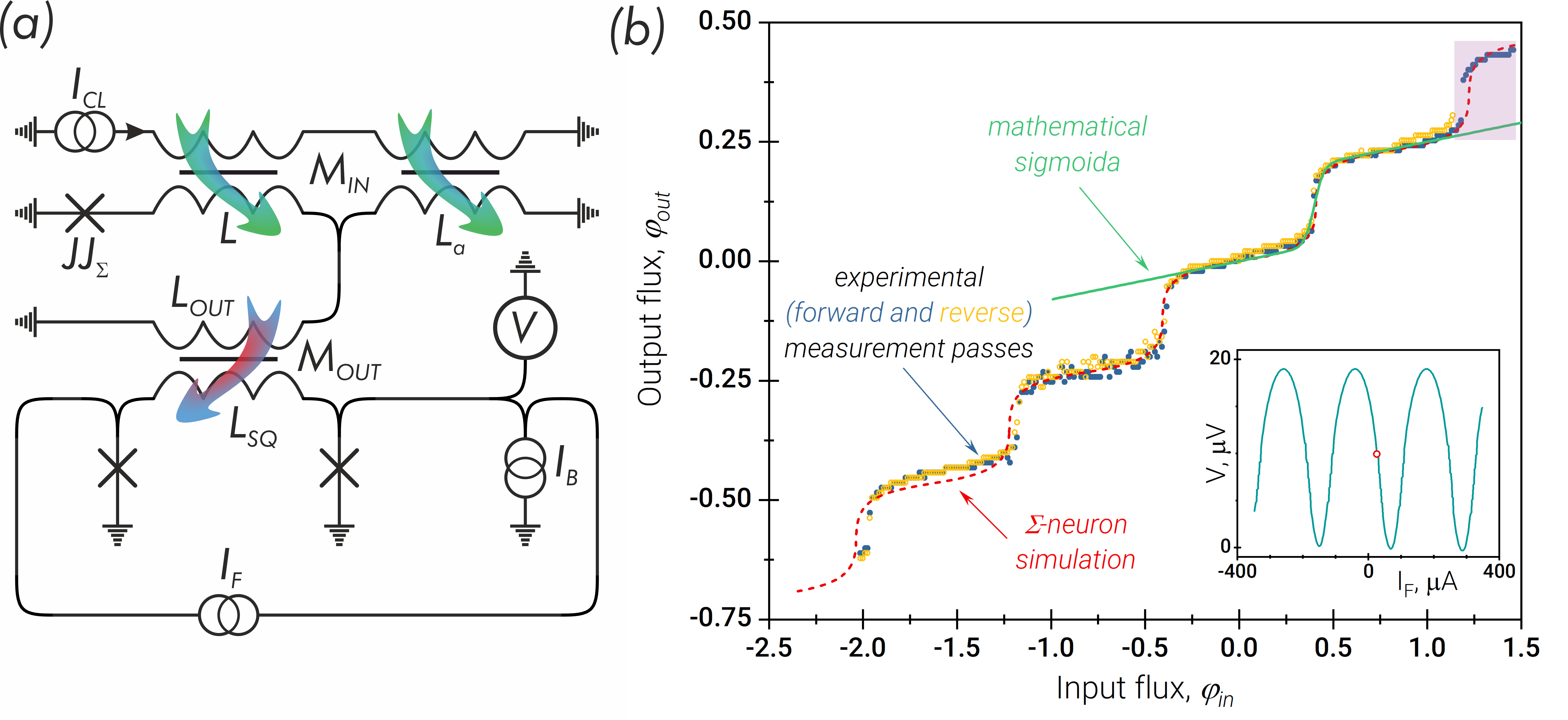}
    \caption{\label{fig_setup_curves}(a) Schematic presentation of the $\Sigma$-neuron, input (control) signal line and readout circuit. (b) Transfer characteristic of the $\Sigma$-neuron: dark-blue and yellow dots -- experimental data of forward and reverse measurements respectively, red dash line -- $\Sigma$-neuron simulation with fitted parameters and green line -- mathematical approximation of experimental data with a sigmoid function. The graph represents the experimental dependence of $(I_{F}-I_{F0})$ on $I_{CL}$ with axes normalised by the values obtained in the experiment and theoretical simulation: $\phi_{in}=I_{CL}M_{in}/\Phi_{0}$, $\varphi_{out}=(I_{F}-I_{F0})L_{out}L_{SQ}/(M_{out}\Phi_{0})$. The mathematical sigmoid is given by a formula of the form $\varphi_{out}(\varphi_{in})=0.08\varphi_{in} + \frac{0.17}{1+exp(-40(\varphi_{in}-0.4))}$. The inset shows the voltage response to the current $I_{F}$ of the readout SQUID -- $IV$-curve. Red dot on it corresponding to the parameters of the initial working point: $I_{B} = 19~\mu\mathrm{A}$, $I_{F0} = 25~\mu\mathrm{A}$, $V_{fix} = 11~\mu\mathrm{V}$.
   The pale pink rectangle indicates the boundaries of the experimental data, within which the data points on the retrapping branch were excluded from the analysis
            }    
\end{figure*}

Previously, in \cite{ionin2023experimental}, a $\Sigma$-neuron sample was experimentally manufactured and measured for the first time, but the results obtained did not yield sufficiently encouraging results. In the present work, we redesigned the $\Sigma$-neuron circuit and more accurately reproduced the parameters from the model at the hardware level. Figure \ref{fig_experiment_photo} shows micrographs of the fabricated $\Sigma$-neuron with control line and readout SQUID, schematically demonstrated in the Fig. \ref{fig_setup_curves}a.

In order to experimentally measure the transfer characteristic of the $\Sigma$-neuron, the last one was used as a nonlinear flux converter, connected magnetically to the gap between the control line ($I_{CL}$) and the readout two-junction SQUID. The readout circuit, in addition to the two-junction SQUID, contains two DC sources ($I_B$ and $I_F$) and a voltmeter ($V$), with the bias current source $I_B$ connected in parallel to one of the Josephson contacts of the SQUID and should not be lower than the maximum non-dissipative current in it ($I_{B}>=I_{MAX}^{SQ}=2I_{C}^{SQ}$), where $I_{C}^{SQ}$ is a critical current of each Josephson junction in the readout SQUID. Compliance with this condition allows us to measure the voltage generated at the readout SQUID and thus control the magnetic flux through the SQUID inductive loop. While current sources $I_{B}$ and $I_{CL}$ and the voltmeter have a common ground, current source $I_{F}$ is connected in parallel with SQUID inductance $L_{SQ}$ and is not connected to the chip's grounding shield.

Before starting measurements, a working point was selected, the parameters of which were chosen based on the ability to operate in the maximum possible dynamic range (red dot in the inset figure \ref{fig_setup_curves}b): $I_{B} = 19\mu\mathrm{A}$ (was selected to obtain the maximum modulation range of the readout SQUID voltage response to the current $I_{F}$), $I_{F0} = 25~\mu\mathrm{A}$ (middle of the $IV$-curve slope of the readout SQUID), corresponding to the $V_{fix} = 11~\mu\mathrm{V}$. When current $I_{CL}$ changes, the operating point on the $IV$-curve of the readout SQUID shifts. This shift was compensated by current $I_{F}$ to return the operating point to its initial state. Thus, the experimentally measured magnetic flux characteristic of the $\Sigma$-neuron represents the dependence of the $(I_{F} - I_{F0})$ on the current $I_{CL}$ in the input (supply) line (Fig. \ref{fig_setup_curves}b). The compensating current $I_{F}$ was determined by solving the equation $f(I_{F})=V(I_{F}) - V_{fix}=0$ using the bisection method. The numerical tolerance was set to $|\Delta f(I_{F})| < 2~\mathrm{pA}$. However, the actual accuracy is limited by noise from the voltmeter, wires, and commutation box, as well as by the current source resolution and SQUID stability. The experimental curve was recorded in both sweep directions (dark-blue and yellow dots in Fig. \ref{fig_setup_curves}b). The forward and reverse sweeps practically coincide over the working range, so the measured transfer characteristic exhibits no appreciable hysteresis and is effectively single-valued. Some points of the retrapping branch of the experimental curve in the right half-plane were removed (the segment inside the pale pink rectangle in Fig.~\ref{fig_setup_curves}b). This is caused by noise and by switching of the voltage operating point to an adjacent SQUID period in $V(I_{F})$ characteristic, and it is a limitation of the measurement setup, not of the $\Sigma$-neuron itself.

 For convenience, the axes in the Fig. \ref{fig_setup_curves}b were converted to normalised magnetic flux according to the values obtained during experimental measurements and theoretical simulations. The inductance $L_{SQ}$ was obtained from measurements of the modulation $I_{B}V$ curves from the field generated by the current $I_{F}$. Mutual inductance $M_{out}$ was calculated from measurements of the $I_{B}V$ curves of reference SQUID structure, where the control line is inductively connected to the SQUID, and the line width and contact geometry were chosen to reproduce the $L_{out}$ of the $\Sigma$-neuron.
 The shunt resistance $R_{N}$ of the Josephson junction incorporated to the $\Sigma$-neuron circuit was obtained by measuring a reference sample with identical parameters. The shunt resistance $R_{N}^{SQ}$ of the readout SQUID and the maximum non-dissipative current value $I_{max}^{SQ}$ were determined by measuring and approximating the $I_{B}V$ curve of the last one in the circuit with the neuron. As a result, the experimentally measured value of $I_{max}^{SQ}$ was an order of magnitude lower than the value specified at the design stage, which is most likely associated with both the size of the Josephson contacts themselves during manufacture process and the noise generated by the $\Sigma$-neuron acting as an RF antenna.

Using a previously developed mathematical model of a $\Sigma$-neuron, we were able to select the necessary parameters and obtain a model transfer characteristic that closely approximates the curve obtained in the experiment (red dashed curve in Fig. \ref{fig_setup_curves}b). To select the parameters, we used the values obtained during the experiment, as well as the inductance values calculated in 3D-MLSI \cite{khapaev20043d, khapaev2015inductance, tarasova2026capabilities}. A comparison of the experimental and model characteristics yielded a Root Mean Square Error (RMSE) estimate of approximately~0.019. The mathematical sigmoid (green line in Fig. \ref{fig_setup_curves}b), constructed to approximate one period of the experimental transfer characteristic of the $\Sigma$-neuron and having the form $y=k_0x + (1 + exp(-k_1(x-x_0))) ^{-1}$, achieved an RMSE value of approximately 0.013. 

\begin{table*}
    \caption{\label{tab_Params} Design, measured and fitted parameters used for modelling the fabricated $\Sigma$-neuron}
    \begin{ruledtabular}
    \begin{tabular}{ c | cc | cc | cccccc }
        \multirow{2}{*}{Values source} & $I_C$ & $I_{\max}^{\mathrm{SQ}}$ & $R_N$ & $R_N^{\mathrm{SQ}}$ 
        & $L$ & $L_a$ & $L_{\text{out}}$ & $L_{\mathrm{SQ}}$ & $M_{\text{in}}$ & $M_{\text{out}}$ \\
        \cline{2-5} \cline{6-11}
          & \multicolumn{2}{ c |}{$\mu\mathrm{A}$} & \multicolumn{2}{ c |}{$\Omega$} & \multicolumn{6}{ c }{\textit{pH}} \\
         \hline\hline
         Design & 50 & 150 & 1.37 & 2.04 & 7.14 & 13.30 & 5.11 & 7.38 & 1.6 & 2.60 \\
         Fitted experiment & (30) & 14 & 1.36 & 2.12 & (7.14) & (13.30) & (5.11) & 9.61 & (1.6) & 2.25 \\
    \end{tabular}
    \end{ruledtabular}
\end{table*}

\begin{figure*}
    \includegraphics[width=0.95\linewidth]{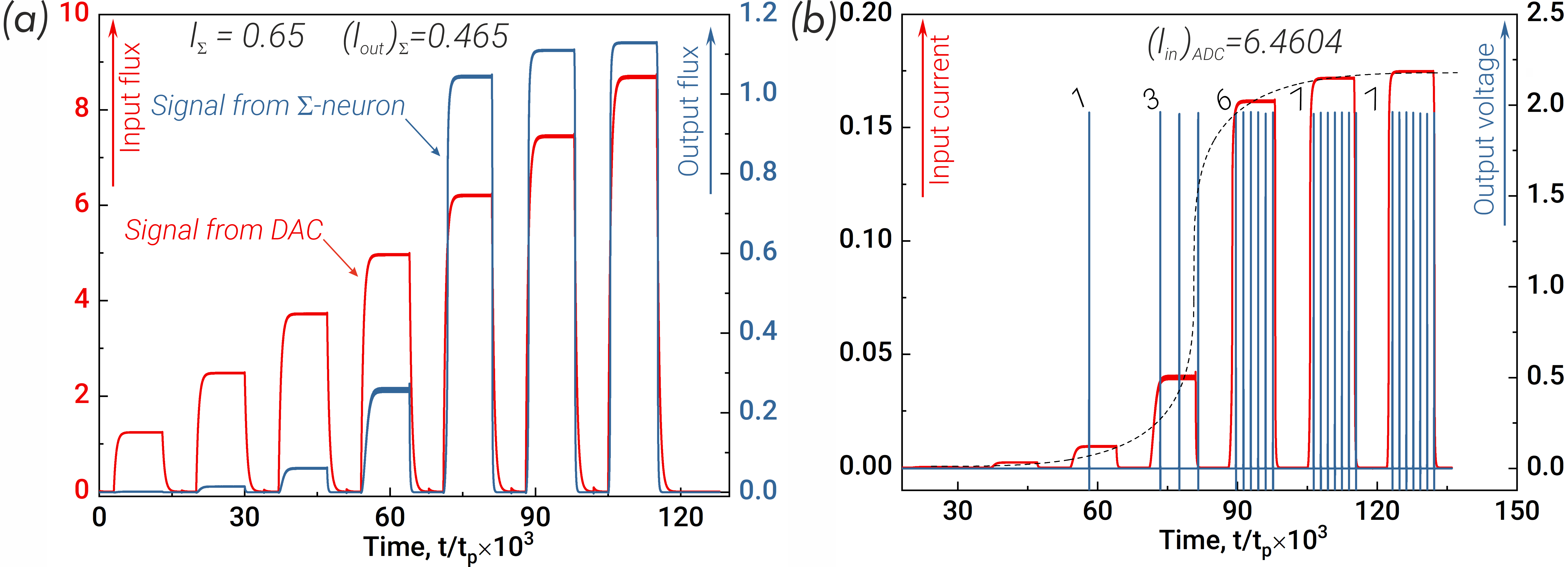}
    \caption{\label{fig_pulses_experiment}(a) Modelling the conversion of the $\{0,1,2,3,4,5,6,7\}$ SFQ signal by 3-bit DAC (red curves) and its conversion through the fitted (model) transfer function of the fabricated $\Sigma$-neuron (dark-blue curves). (b) Simulation ADC voltage response (dark-blue curves) after conversion of the signal, generated by the fitted (model) transfer function of the fabricated $\Sigma$-neuron (red curves). Parameters of the fitted $\Sigma$-neuron are: $l = 0.65$, $l_{out} = 0.465$, $l_a = 1 + l$. The ADC parameters are the same as those specified in the Fig.~\ref{fig_pulses} caption, with the exception of $L_{in}/L_0=6.4604$, $K_{\rm out}=0.155$.
            }    
\end{figure*}

For the end-to-end simulation we use a model transfer characteristic fitted to the measured data (Fig.~\ref{fig_sigma_transfer}), in which the additional inductance is set to its target value $l_a = 1.65$ (18.1~pH) rather than the $l_a = 1.211$ (13.3~pH) extracted from the fabricated device (Table~\ref{tab_Params}). This is not a cosmetic reshaping of the curve: $l_a = 1.65$ is the design target at which the horizontal $\Sigma$-shelves of the transfer characteristic are properly developed, and a corrected version of the neuron is expected to reach it while keeping the other present parameters. In the fabricated sample this inductance came out lower, which weakens the saturation, so simulating at $l_a = 1.65$ represents the intended, redesigned neuron rather than an arbitrary adjustment made for comparison. 
This fitted (model) characteristic, constrained by the measured data, has a distinct non-linearity whose effect on the input-signal transformation is significant. Figure~\ref{fig_pulses_experiment} shows a digital-to-analog-to-digital conversion simulation that uses this fitted characteristic (shown in Fig.~\ref{fig_setup_curves}b), performed in the same way as for the model-optimised $\Sigma$-neuron characteristic (Fig.~\ref{fig_pulses}).

As expected, the output of the fitted (model) $\Sigma$-neuron characteristic becomes markedly nonlinear, and the compression of its upper branch is visible directly in the number of output spikes. With the ADC settings kept fixed (despite the value of ADC input inductance which after normalisation equal to $(l_{in})_{ADC}=6.4604$), the supra-threshold inputs are mapped, in order, to $1$, $3$, $6$, $7$ and $7$ output spikes (Fig.~\ref{fig_pulses_experiment}b). The two largest inputs, $\{6\}$ and $\{7\}$, both saturate to $7$ and therefore become indistinguishable at the output, whereas the smaller inputs give only a few spikes. All counts remain within the $0$--$7$ range of the 3-bit converter. This follows from the change in the magnitude range of the transfer characteristic: the larger inputs shift onto its saturating upper branch, while the smaller inputs stay on the lower, near-linear part, as in the model-optimised conversion of Fig.~\ref{fig_pulses}.

%%%%%%%%%%%%%%%%%%%%%%%%%%%
\section{\label{Sec:MNIST}Simulation of a Neural Network Based on $\Sigma$-Neurons}
As a test problem we use the classification of handwritten digits from the MNIST dataset \cite{lecun1998mnist}. MNIST is one of the most widely used reference benchmarks for new network architectures: the input images have a fixed size, the labelled set is large, and the results can be compared directly with the published literature.

We solve the task with a multilayer perceptron with a single hidden layer. Although convolutional networks dominate image recognition, the perceptron is the more informative choice for a superconducting neural processor: its simple structure lets each artificial neuron be mapped onto a separate hardware element, so the influence of the neuron's physical characteristics on classification can be studied without the additional effects introduced by convolution. A single hidden layer of $300$ to $500$ neurons reaches about $97$--$98\%$ accuracy on MNIST, while adding neurons beyond this range brings only a small further gain at a large increase in cost \cite{lecun1998mnist}. Architectures of this class are therefore a common compromise between classification quality and hardware complexity.

Two neural networks with identical architectures were developed. Each network maps the $28\times28$ input pixels of MNIST images through a fully connected layer of $N=500$ hidden units, each applying the $\Sigma$-neuron transfer function, to a linear read-out with ten class outputs. Both linear layers omit bias terms. A batch-normalisation stage \cite{ioffe2015batchnorm} precedes the hidden activation and the read-out, and a rectified-linear read-out with an $\arg\max$ decision assigns the class. Training minimises the mean-squared error between the read-out and the one-hot label. The first network used an ideal logistic sigmoid, while the second used a look-up table (LUT) of the measured $\Sigma$-neuron transfer characteristic, evaluated by linear interpolation (this characteristic is shown in Fig.~\ref{fig_setup_curves}b). To ensure the proper functioning of the network, a single period (approximately from $-0.2$ to $1$ on the input flux) was selected from the entire experimental curve shown in Fig. \ref{fig_setup_curves}b as blue points. The linear component, determined from the approximation and equal to $0.08\varphi_{in}$, was then subtracted from it. Next, the range of values for this period was normalised to one, and the domain (dynamic range) was centred at zero.

For the training algorithm we followed current practice. Resilient backpropagation (RProp) was long used for multilayer perceptrons and performs well for small networks trained in full-batch mode \cite{Riedmiller1993}, but the move to mini-batch learning, the spread of ReLU activations, and the growth in the number of trainable parameters have shifted the field to adaptive optimisers of the Adam family. Adam combines momentum with RMSProp and sets an individual learning rate for each weight from running estimates of the first and second moments of the gradient \cite{kingma2015adam}. This gives fast, noise-tolerant convergence with little manual tuning of the learning rate, and it is now the de~facto standard for training perceptrons on image-classification tasks. As a regulariser we use dropout \cite{srivastava2014dropout}, which randomly removes a fraction of the hidden units during training. A rate of $0.2$ reduces the gap between training and test accuracy without a noticeable increase in cost.

The remaining design elements are dictated by the hardware that the network is intended to emulate. The weights, which correspond to physical couplings of bounded strength, are restricted to the range $[-1/N,\,+1/N]$ and are hard-clipped to this interval after every update. The $1/N$ scale keeps the weighted sum reaching each neuron within the input-flux window over which the $\Sigma$-neuron transfer characteristic is defined, so the analog cell is never driven outside its operating range. Because such tight bounds leave the optimiser little room to move, the batch-normalisation stage works as an operating-point control: it recentres and rescales the accumulated input onto the sensitive part of the transfer characteristic (in our case at $\varphi_{in}=0$), in the same way that a bias current sets the working point of the fabricated device. For the same reason the learning rate starts relatively high ($0.005$) and is annealed towards $10^{-4}$ on a cosine schedule over $20$ epochs, so the larger early steps let the bounded weights reorganise while the decaying rate allows fine adjustment within the clipped range. The decision to use exactly 20 training epochs was based on the minimum requirement for recognition accuracy to exceed 90\%. An attempt to increase the number of training epochs to 50 resulted in a negligible improvement in recognition accuracy at the second decimal place.

\begin{figure}
    \centering
    \includegraphics[width=0.95\linewidth]{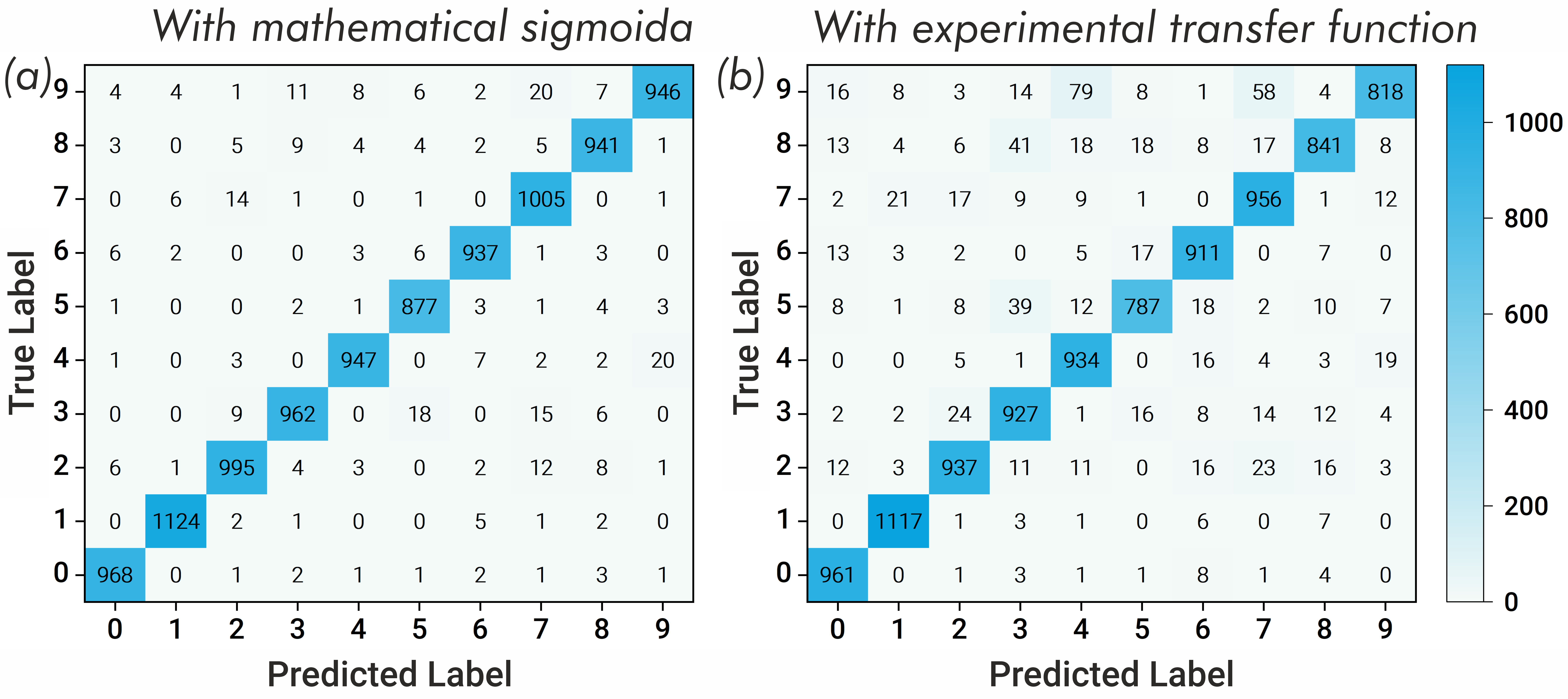}
    \caption{\label{fig:confmat}Confusion matrices on the MNIST test dataset ($10^{4}$ images) for the two perceptron networks: (a)~Math-based perceptron, with the ideal logistic sigmoid as the hidden activation, and (b)~Experimental-based perceptron, with the experimentally measured $\Sigma$-neuron transfer characteristic used as a look-up table. The colour encodes the number of test images
            }
\end{figure}

In summary, we trained two separate networks with the same architecture ($784$ inputs, one hidden layer of $N=500$ units, ten outputs) and the same optimisation protocol, differing only in the hidden activation. In the first network, denoted as Math-based perceptron (MBP), the hidden units use an ideal logistic sigmoid, while in the second, Experimental-based perceptron (EBP), they use the experimentally measured $\Sigma$-neuron transfer characteristic supplied as a LUT. On the MNIST test dataset the two networks reach a classification accuracy of $97.0\%$ (macro-averaged $F_1 = 0.970$) and $91.9\%$ (macro-averaged $F_1 = 0.918$), respectively. The corresponding confusion matrices are shown in the Fig.~\ref{fig:confmat}.

Taken together, replacing the idealised activation by the measured $\Sigma$-neuron characteristic costs about five percentage points of accuracy, from $97.0\%$ to $91.9\%$, while leaving the network fully trainable and free of overfitting. The cost is modest and physically interpretable, which supports the view that the fabricated $\Sigma$-neuron is usable as an activation element in a perceptron-like classifier, and that the main route to closing the remaining gap is to widen the usable dynamic range of the neuron and to adapt the training procedure to the device. In the Tab. \ref{tab:NN_compare} we have compiled data on key metrics to compare the performance of the two perceptron networks developed.

\begin{table}[t]
\caption{\label{tab:NN_compare}Comparison of the two perceptron networks on the MNIST test dataset ($10^{4}$ images). Both networks share the architecture and training protocol (no bias, batch normalisation, dropout $0.2$, rectified-linear read-out, MSE loss, Adam with a cosine schedule from $0.005$ to $10^{-4}$, $20$ learning epochs, weights clipped to $[-1/500,\,+1/500]$) and differ only in the hidden activation.
        }
\begin{ruledtabular}
\begin{tabular}{lcc}
Metric & Math-based perceptron & Experimental-based perceptron \\
\colrule
Hidden activation function   & ideal sigmoid & measured $\Sigma$-neuron LUT \\
Test accuracy                & $97.0\%$      & $91.9\%$ \\
Training accuracy            & $96.35\%$     & $89.42\%$ \\
Macro-averaged $F_1$         & $0.970$       & $0.918$ \\
Best per-class $F_1$         & $0.989$ (digit 1) & $0.974$ (digit 1) \\
Worst per-class $F_1$        & $0.955$ (digit 9) & $0.870$ (digit 9) \\
Largest confusion\footnote{Largest off-diagonal count in the confusion matrix,
given as true$\rightarrow$predicted.}
                             & $4\!\rightarrow\!9$ ($20$) & $9\!\rightarrow\!4$ ($79$) \\
\end{tabular}
\end{ruledtabular}
\end{table}

%%%%%%%%%%%%%%%%%%%%%%%%%%%
\section{Discussion}
The central outcome of this work is a proof-of-principle for a DAD converter tailored to superconducting perceptron-like neural networks, where digital SFQ pulse representations are converted into analog levels for high-precision non-linear processing and then mapped back into SFQ form for subsequent logic and routing. In the proposed hybrid concept (illustrated as the “DAC + $\Sigma$-neuron + ADC” chain), the analog $\Sigma$-neuron serves as an activation-function block, while the surrounding SFQ-compatible converters provide programmability and composability with larger superconducting digital systems. Importantly, the modelling flow is anchored to experimentally obtained $\Sigma$-neuron characteristics, linking circuit-level feasibility to realistic device parameters rather than idealised assumptions. Together, these elements clarify how a mixed-domain architecture can mitigate well-known limitations of purely analog (calibration, variability, noise sensitivity) and purely digital (resource-intensive high-precision arithmetic) superconducting neural implementations while preserving the main promise of superconducting neuromorphic hardware: high-speed operation at ultra-low dissipation.

From the digital-to-analog side, the presented DAC architecture implements amplitude quantisation by combining SFQ logic elements (D flip-flops acting as programmable gates) with weighted SQUID-based voltage sources whose contributions sum to the output level. The 3-bit demonstrator schematics make explicit how binary weighting is realised physically by assigning an increasing number of identical converting elements per bit (Fig. \ref{fig_DAC_ADC}a), and the time-domain examples (Fig. \ref{fig_pulses}a) illustrate how the digitally encoded signal is converted into an analog level, which is then processed by the $\Sigma$-neuron through its activation function. Conceptually, this is attractive because it keeps the information-bearing representation digital -- robust SFQ pulse trains -- transferring only the calculation of the activation function value to the analog domain.
At the same time, the simulation results highlight a practical limitation of this approach: the transition from digital representation to analog must take into account relaxation processes in the analog neuron circuit. Neglecting these processes can result in an incorrect and unreproducible signal at the system output. Therefore, the accuracy of signal conversion and the throughput of the neuron depend not only on the synchronisation of the SFQ at the DAC input, but also on the filtering/averaging strategy and impedance matching between the DAC output and the neuron input.

The $\Sigma$-neuron operation results emphasise why an analog activation block remains valuable even in a primarily digital superconducting network. The neuron's transfer characteristic performs nonlinear compression of the input signal with saturation, limiting the output range to approximately from 0 to 1 and suppressing large input signal values. Reproducing this functionality digitally with comparable precision would require a significant increase in circuit complexity. A key methodological point is that the $\Sigma$-neuron behaviour used in the end-to-end simulations is not treated as an abstract nonlinearity, but is instead derived from fitted parameters based on experimental characterisation (Table \ref{tab_Params}). This closes the loop between the physical device and system-level simulation: it becomes clear which aspects of the neuron activation function are robust to manufacturing variations and which require parameter tuning, bias current optimisation, or even redesign to ensure the desired sigmoidal transformation in large superconducting digital-to-analog circuits.

On the analog-to-digital side, the ADC (Fig. \ref{fig_DAC_ADC}b) is shown in simulation to translate the $\Sigma$-neuron output waveform into SFQ pulse sequences through an integrating front-end and Josephson switching dynamics, so that pulse generation rate and timing depend on both the level and the temporal evolution of the neuron signal (Fig.~\ref{fig_pulses}b). In the chosen operating regime, the ADC was configured so that the number of generated SFQ pulses on its output corresponds to the converted decimal numerical value, enabling a direct demonstration of functional end-to-end processing through the chain (“DAC -- $\Sigma$-neuron -- ADC”). Simulation of the chain with different $\Sigma$-neuron transfer characteristics clearly demonstrated the importance of matching the DAC and ADC parameters in the process of conversion of the same signal: while maintaining the parameters of the converters, replacing the transfer characteristics of the central element ($\Sigma$-neuron) leads to significant changes in the number of output spikes, which can be misleading.

We emphasise that the present study is deliberately block-level: the DAC, the $\Sigma$-neuron and the ADC were each simulated separately, and the complete DAC--$\Sigma$-neuron--ADC chain was not co-simulated as a single self-consistent circuit. The blocks are linked through explicit interface quantities rather than through a shared electrical network. The DAC output sets the drive level of the neuron, and the neuron output current $i_{\rm out}(t)$ enters the ADC through $i_{\rm ADC}=i_0+K_{\rm out}\,i_{\rm out}$ (Eq.~(\ref{eq:ADCinput})), where $K_{\rm out}$ is treated as a calibrated matching coefficient. Within this description we assume matched normalisation between stages and neglect loading and impedance effects, the back-action of the ADC on the neuron, and the detailed dynamics of the DAC output network and its low-pass filtering, which enter only through their averaged effect. A self-consistent simulation of the integrated chain, in which the DAC output network, the voltage-to-flux/current conversion, the coupling geometry and the ADC input impedance are modelled together with their mutual back-action, would require a dedicated study of interface optimisation and is the subject of our future work.

The simulation results presented for input data relevant to SFQ-based operation clearly demonstrate how neuron nonlinearity can be incorporated into an information flow compatible with SFQ logic without requiring the construction of a digital neural block that emulates the sigmoid activation function.
A practical limitation of the present design is that the $\Sigma$-neuron, originally optimised for adiabatic operation, raises the per-symbol processing time to the order of tens of nanoseconds, during which the ADC remains in the resistive state. As a result, although the neuron itself dissipates only fractions of an attojoule, the dissipation of the whole structure rises into the femtojoule range. To illustrate this, we estimated the energy dissipation for the circuit under consideration. Because of the DAC's architecture, where the output signal is generated by a cascade of SQUIDs (see Fig.~\ref{fig_DAC_ADC}), it is not quite accurate to talk about dissipation per bit of information. Therefore, we estimated the amount of energy dissipated when processing signal $\{1\}$ (001) -- given the minimum possible dissipation, and signal $\{7\}$ (111) -- given the maximum possible dissipation, for a 3-bit bus. For simplicity, we assume that the critical current values for all Josephson junctions in the chain are identical and equal to $I_C=30~\mu A$ (the critical current value obtained from experimental measurements of a $\Sigma$-neuron). The results of this evaluation are shown in the Tab.~\ref{tab:dissipation}. For the selected critical current value, the normalisation energy $E_C = 0.01~aJ$ ($9.88\times 10^{-21}~J$) and the normalisation time $t_p = 11~ps$. Then the $\Sigma$-neuron releases an average of $35~yJ$ to $3.61~zJ$, while the ADC is of the order of $65~zJ$ to $0.45~aJ$. The ''hottest'' element in the circuit remains the DAC, which, in order to generate sufficiently long pulses ($\sim 11\,000~t_p$), is forced to keep the cascade of SQUIDs in the resistive mode, resulting in energy dissipation of the order of $74$ to $520~aJ$. However, nearly one-third of all the energy dissipated by the DAC is accounted for by the low-pass RC filter. Even though reducing the pulse duration is the simplest way to lower this value, we would like to emphasise once again that the current circuit configuration is intended solely as a demonstration platform. Future research will focus on exploring the possibility of using more energy-efficient interfaces and converters. The symbol-processing time -- whether signal $\{1\}$ or $\{7\}$ -- is determined by the adiabatic operating mode of the $\Sigma$-neuron and, in simulations, was on the order of $120~ns$.

\begin{table}
\caption{\label{tab:dissipation}Estimations of energy dissipation in DAC, $\Sigma$-neuron and ADC in a chain for using theoretically ideal and fitted from the experiment $\Sigma$-neuron transfer characteristics
        }
\begin{ruledtabular}
\begin{tabular}{lcccc}
Metric & Digit & DAC & $\Sigma$-neuron (ideal/fitted) & ADC (ideal/fitted) \\
\colrule
\multirow{2}{*}{Dissipated energy, $E_C$}    & 1 (001) &  7 500\footnote{$\sim 5~500~E_C$ for a cascade of SQUIDs and $\sim 2~000~E_C$ for a low-pass RC filter.} &  0.0047 / 0.0023 & 6.49 / 6.65\\

                                             & 7 (111) &  52 500 &  0.35 / 0.38 & 45.45 / 46.55\\
\colrule
\multirow{2}{*}{Symbol-processing time, $t_p$} & 1 (001) & \multicolumn{3}{c}{\multirow{2}{*}{$\sim 11~000$}} \\
                                          & 7 (111) & \multicolumn{3}{c}{} \\
\end{tabular}
\end{ruledtabular}
\end{table}

Beyond the single-symbol conversion, the network-level test places the measured $\Sigma$-neuron in a realistic setting. Using the fabricated-device transfer characteristic as the hidden activation of a single-hidden-layer perceptron, and constraining the weights to the bounded, clipped range accessible to physical couplings, the network classifies MNIST digits with $91.9\%$ accuracy, against $97.0\%$ for the same network with an ideal sigmoid. The gap of about five percentage points is modest and physically interpretable: the reduced output swing and the steep, saturating shelves of the measured characteristic compress the separation between visually similar digits, so that the additional errors concentrate on a few confusable pairs (for example $9\rightarrow4$ and $8\rightarrow3$) rather than degrading all classes uniformly. Two conclusions follow. First, the fabricated $\Sigma$-neuron works as an activation element at the network level, not only as an isolated conversion cell, which is the strongest evidence so far that the measured nonlinearity supports learning. Second, the route to closing the remaining gap is to widen the usable dynamic range of the neuron and to adapt the learning process to the device, rather than to change the classifier itself. This links the present block-level study to hardware-aware (physics-aware) learning, in which the measured device response is built into the forward pass.

The observations outlined above highlight several distinct directions for the further development of the presented approach.
First, converter–neuron co-design should be treated as a primary objective: increasing the neuron bandwidth (or redesigning the operating point away from strongly adiabatic dynamics) and shortening the time the ADC spends in dissipative regimes are likely to yield the largest gains in throughput-per-energy. Second, scaling resolution beyond the demonstrated few-bit regime will require careful architectural choices, since straightforward binary weighting increases hardware cost rapidly; segmented or time-multiplexed conversion strategies may offer better system-level scaling while preserving SFQ robustness. Third, the experimentally observed parameter dispersion implied by the need for fitting underscores the importance of biasing strategies, calibration protocols, and variability-aware design rules if large arrays of $\Sigma$-neurons are to be deployed. Finally, a natural next experimental milestone is an integrated measurement of the full “DAC + $\Sigma$-neuron + ADC” chain on-chip under representative clocking, establishing not only static transfer properties but also noise margins, timing tolerances, and the energy breakdown across the blocks. These metrics will ultimately determine viability for multilayer superconducting perceptron-like networks.

%%%%%%%%%%%%%%%%%%%%
\section{Methods}
\textbf{$\Sigma$-neuron fabrication technology.} A 100 nm layer of $Al_{2}O_{3}$ is deposited onto a silicon substrate by RF magnetron sputtering as a stop layer for subsequent plasma-chemical etching. The trilayer structure $Nb-AlO_{x}-Nb$ is formed in a single vacuum cycle in an ultra-high vacuum deposition system with a base pressure of $10^{-8}~mbar$, equipped with cryogenic and turbomolecular pumps. The thickness of the bottom niobium layer is $200~nm$, followed by deposition of $7~nm$ of aluminium, which is then oxidised at a pressure of $1.5~mbar$ for 20 minutes. After that, the top niobium layer with a thickness of $80~nm$ is deposited.
The geometry of the structures on the sample is defined using photolithography with photomasks fabricated by electron-beam lithography. Superconductor/insulator/superconductor (SIS) Josephson junctions of the required area are formed using the SNEAP (Selective Niobium Etching and Anodization Process) method, in which the top niobium layer is removed by reactive ion etching in a $CF_{4}$ atmosphere, with the aluminium layer serving as the etch stop. Then, the sidewalls of the SIS junctions are anodised in an electrolyte solution to eliminate possible micro-shorts. After that, a $SiO_{2}$ insulation layer is deposited to provide electrical isolation between the base and top wiring electrodes. In the next deposition vacuum cycle, the top electrode layer of niobium with a thickness of $350~nm$ is formed.

\textbf{Measurement setup.} All measurements were carried out in a closed-cycle Gifford–McMahon refrigerator at 2.9 K. The $\Sigma$-neuron was enclosed in a copper sample box shielded from non-equilibrium infrared photons coming from the 2.9 K stage. Measurement of the $I$–$V$ curve was carried out by the 4-point (Kelvin) method using separate twisted pairs for current and voltage contacts of the structures. The current was passed to the $\Sigma$-neuron, and the voltage was measured via twisted-pair beryllium bronze lines filtered with a low-pass lumped-element RC-filter ($R = 100~\Omega$, $C = 100$ nF, cut-off frequency $f = \frac{1}{2\pi RC} = 15.9~\text{kHz}$ at the 2.9 K stage \cite{nazhestkin2025enhancing}). 
Current measurements were taken using a Keithley 6221 precision current source, whilst voltage measurements were taken using a Keithley 2182A nanovoltmeter with a signal integration time of 5 power line cycles (PLC).

%%%%%%%%%%%%%%%%%%%%
\section{Conclusion}
In this work, we proposed and investigated a hybrid digital–analog–digital converter for superconducting perceptron-like neural networks based on a DAC, a $\Sigma$-neuron and an ADC. Numerical simulations show that SFQ pulse sequences can be converted into analog signal levels, non-linearly transformed by the $\Sigma$-neuron, and then mapped back into SFQ form. As a key experimental step, we fabricated and characterised a redesigned $\Sigma$-neuron and obtained a sigmoid-like transfer characteristic in reasonable agreement with the fitted model. By incorporating experimentally extracted neuron parameters into system-level simulations, we demonstrated how the transfer characteristic of a real device affects end-to-end signal conversion and identified the importance of converter–neuron matching. These results support the feasibility of the proposed hybrid architecture and provide a concrete basis for the next stage of the work: experimental on-chip implementation and testing of the full ''DAC + $\Sigma$-neuron + ADC'' chain or base on the \cite{bastrakova2025digital} ''$\Sigma$-neuron + ADC'' chain with digital control.

The digital parts of both converters were implemented with compact all-JJ DFF and TFF (Fig.~\ref{fig_FF}), and the all-JJ TFF counter maps the analog $\Sigma$-neuron output back into an SFQ pulse count, with the ADC calibration and time-domain operation shown in Fig.~\ref{fig_ADC_Demo}. Finally, using the measured $\Sigma$-neuron transfer characteristic as the hidden activation of a single-hidden-layer perceptron, we trained two networks on the MNIST dataset and obtained test accuracies of $97.0\%$ with an ideal sigmoid and $91.9\%$ with the measured characteristic (Fig.~\ref{fig:confmat} and Tab.~\ref{tab:NN_compare}), which shows that the fabricated neuron remains usable as an activation element at the network level.

\begin{acknowledgments}
The numerical simulations were performed with the financial support of the Russian Science Foundation, Grant No. 24-19-00187.
We acknowledge partial support for chips designs from the Rosatom within the framework of the Roadmap for Quantum computing (Contract No. 868-1.3-15/15-2021 dated October 5).
 
We would like to extend special thanks to the MTUCI Lead Engineer \textit{Y. M. Kupriyanov} for his assistance in writing and debugging the neural network code.
\end{acknowledgments}

\textbf{Code availability.} The source code used to train and evaluate the $\Sigma$-neuron perceptron on MNIST is openly available at https://github.com/aeschegolev/sigma-neuron-mnist.git and archived at Zenodo https://doi.org/10.5281/zenodo.21263269.

\section*{References}
\bibliography{sigma_bibtex}

\end{document}